%% file: main.tex
\documentclass[sigconf,dvipsnames]{acmart}
\usepackage[utf8]{inputenc}
\usepackage[T1]{fontenc}
\usepackage[inline]{enumitem}
\usepackage{float}
\usepackage{booktabs}
\usepackage{amsmath}
\usepackage{amsfonts}
\usepackage{enumitem}
\usepackage{mathtools}
\usepackage{hyperref}
\usepackage{url}
\usepackage{microtype}
\usepackage{natbib}
\usepackage{multirow}
\usepackage{graphicx}
\usepackage{subcaption}
\usepackage{xcolor}
\usepackage{algorithm}
\usepackage{algorithmic}
\usepackage{balance}
\usepackage{titlesec}
\usepackage{amsmath}
\usepackage{newtxmath}
\newcommand{\ie}{\emph{i.e.}}
\newcommand{\eg}{\emph{e.g.}}

\newcommand{\shortto}[1][3pt]{\mathrel{%
   \hbox{\rule[\dimexpr\fontdimen22\textfont2-.2pt\relax]{#1}{.4pt}}%
   \mkern-4mu\hbox{\usefont{U}{lasy}{m}{n}\symbol{41}}}}

\titlespacing*{\section}{0pt}{0.7\baselineskip}{0.3\baselineskip}
\titlespacing*{\subsection}{0pt}{0.5\baselineskip}{0.2\baselineskip}

\makeatletter

\newcommand{\scriptshortto}[1][3pt]{{%
    \hbox{\rule[\scriptratio\dimexpr\fontdimen22\textfont2-.2pt\relax]
               {\scriptratio\dimexpr#1\relax}{\scriptratio\dimexpr.4pt\relax}}%
   \mkern-4mu\hbox{\let\f@size\sf@size\usefont{U}{lasy}{m}{n}\symbol{41}}}}

\makeatother

\AtBeginDocument{%
  \providecommand\BibTeX{{%
    \normalfont B\kern-0.5em{\scshape i\kern-0.25em b}\kern-0.8em\TeX}}}

\copyrightyear{2022}
\acmYear{2022}
\setcopyright{rightsretained}
\acmConference[WWW '22]{Proceedings of the ACM Web Conference 2022}{April 25--29, 2022}{Virtual Event, Lyon, France}
\acmBooktitle{Proceedings of the ACM Web Conference 2022 (WWW '22), April 25--29, 2022, Virtual Event, Lyon, France}\acmDOI{10.1145/3485447.3512262}
\acmISBN{978-1-4503-9096-5/22/04}

\newif\ifanon
\anontrue

\begin{document}
%\fancyhead{}

% \title[Exposing Query Identification as a Retrieval Problem]{Exposing Query Identification as a Retrieval Problem}
\title[Exposing Query Identification for Search Transparency]{Exposing Query Identification for Search Transparency}

\author{Ruohan Li}
\affiliation{%
  \institution{Carnegie Mellon University, Microsoft}
%  \city{city} 
%  \state{state}
   \country{United States}
%   \postcode{zipcode}
}
\email{ruohanli@microsoft.com}

\author{Jianxiang Li}
\affiliation{%
  \institution{Carnegie Mellon University, Microsoft}
%  \city{city} 
%  \state{state}
  \country{United States}
%   \postcode{zipcode}
}
\email{jianxiangli@microsoft.com}

\author{Bhaskar Mitra}
\affiliation{%
  \institution{Microsoft, University College London}
%  \city{city} 
%  \state{state}
   \country{Canada}
%   \postcode{zipcode}
}
\email{bmitra@microsoft.com}

\author{Fernando Diaz}
\affiliation{%
  \institution{Mila Quebec \\ Artificial Intelligence Institute}
%  \city{city} 
%  \state{state}
  \country{Canada}
%   \postcode{zipcode}
}
\email{diazf@acm.org}

\author{Asia J. Biega}
\authornote{Work partly done while at Microsoft Research.}
\affiliation{%
  \institution{Max Planck Institute\\ for Security and Privacy}
%  \city{city} 
%  \state{state}
  \country{Germany}
%   \postcode{zipcode}
}
\email{asia.biega@mpi-sp.org}

\input{01-abstract}



\maketitle

\input{02-intro}
\input{03-related}
\input{04-task}

\input{06-ance}
\input{05-bm25}
\input{07-experiment}
\input{08-result}
\input{09-conclusion}

%\newpage
\balance
\bibliographystyle{ACM-Reference-Format}
\bibliography{main}

\appendix
\input{10-appendix}

\end{document}

%% file: 01-abstract.tex
\begin{abstract}
Search systems control the exposure of ranked content to searchers.
In many cases, creators value not only the exposure of their content but, moreover, an understanding of the specific searches where the content is surfaced.
The problem of identifying which queries expose a given piece of content in the ranked results is an important and relatively underexplored search transparency challenge. Exposing queries are useful for quantifying various issues of search bias, privacy, data protection, security, and search engine optimization.
Exact identification of exposing queries in a given system is computationally expensive, especially in dynamic contexts such as web search. We explore the feasibility of approximate exposing query identification (EQI) as a retrieval task by reversing the role of queries and documents in two classes of search systems: dense dual-encoder models and traditional BM25. We then improve upon this approach through metric learning over the retrieval embedding space. We further derive an evaluation metric to measure the quality of a ranking of exposing queries, as well as conducting an empirical analysis of various practical aspects of approximate EQI. Overall, our work contributes a novel conception of transparency in search systems and computational means of achieving it.
\end{abstract}

%% file: 02-intro.tex
\section{Introduction}
\label{sec:intro}

% What is search exposure
Given a large repository of searchable content, information retrieval (IR) systems control the exposure of the content to searchers. In many cases, creators value not only the exposure of their content but also an understanding of the search contexts in which their content is surfaced. To enable transparency about search contexts facilitated by a given system, search engines should identify which \emph{search queries expose each of the corpus documents in the ranking results}~\cite{biega2017learning}. Yet, search systems currently do not provide such transparency to content producers, as few non-brute-force methods exist that can be used for this task. Identification of such exposing queries is our focus in this paper.

\vspace{.3em}
\noindent\textbf{Exposing Query Use Cases.} Evidence that exposure transparency is important permeates multiple lines of work, including fairness and bias, privacy, search engine optimization, and even security. \citet{azzopardi2008retrievability} defined the \emph{retrievability bias} as the disparity in the weighted sum of ranks different documents are retrieved at over all possible queries. Finding which queries expose which documents is therefore necessary to quantify and audit this bias. Fair rankings typically focus on particular queries~\cite{diaz2020evaluating}, while there is virtually no way for an individual user to get an overview of which rankings overall expose their content. This is especially crucial in settings where exposure in rankings is highly monetizable, like hiring~\cite{geyik2019fairness} or online marketplaces~\cite{mehrotra2018towards}. \citet{biega2017learning} argued that exposing queries are an important \emph{privacy} tool in search systems, demonstrating a range of scenarios where \emph{exposure to sensitive queries} is problematic~\cite{peddinti2014cloak}. The \emph{right to be forgotten}~\cite{bertram2019five} is one of the key data protection rights: Knowledge of exposing queries might enable users to execute this right more precisely than removing themselves from search results completely. 
Content creators could furthermore use the exposure transparency to aid in improving the performance of their documents in search results~\cite{raifer2017information,vasilisky2020studying}.
Beyond content creators, service providers may also care about understanding how different content is exposed~\cite{otterbacher2017competent, wilkie2017algorithmic} to better understand their systems.
Finally, controlling which queries expose certain types of content might prevent unintended leaking of information through search engines
~\cite{fbLeak}.

\vspace{.3em}
\noindent\textbf{Approaches.} An intuitive strategy for identifying exposing queries is to analyze past search logs to find which documents were actually returned in response to specific queries.
Yet, such logs quickly lose their relevance as a source of exposing queries as the search ecosystem is constantly changing and both the collection to be searched and the search models are updated.
Furthermore, in certain sensitive search contexts, the potential for exposure should ideally be captured before exposure occurs and is logged. Thus, exposing queries need to be computed based on a given search ranker rather than from logs. Exact brute-force computation assuming we have a universe of feasible queries---for instance, based on existing query logs and/or generated from the corpus documents---is practically inefficient as it involves issuing all these queries to the retrieval system and re-aggregating the ranking results. The question then becomes of whether we can do better.

\vspace{.3em}
\noindent\textbf{Contributions.}
In this paper, we explore the feasibility of approximate exposing query identification (EQI) as a \emph{retrieval} task in which the function of queries and documents is \emph{reversed}.
We study this idea in the context of two classes of search systems: dense embedding-based nearest neighbor search---\eg,~\cite{huang2013learning, nalisnick2016improving, xiong2020approximate} and traditional BM25-based~\cite{robertson2009probabilistic} retrieval using inverted-index~\citep{zobel2006inverted}.
In case of embedding-based search systems, we show how retrieval-based EQI can be improved if we treat the problem as a form of metric learning in a joint query-document embedding space, transforming an embedding space of a document retrieval model such that a nearest neighbor search in the transformed embedding space approximately corresponds to a reverse nearest neighbor search over the original embeddings.
In search systems, whether a query exposes a document is a function of both the document relevance as well as how many other documents in the corpus are relevant.
Our intuition, therefore, is that we might be able to reorganize queries and documents in the embedding space by adequately selecting training data that captures the corpus characteristics of dense document regions.

We furthermore derive an evaluation metric to measure the quality of a ranking of exposing queries similar to two-sided metrics previously used for assessing the effectiveness of query suggestion algorithms~\cite{santos2013learning}.
The metric reveals that the exposure significance of a query is dependent on the behavior of both the document retrieval users and the EQI system user.
Overall, our work contributes a novel conception of transparency in search systems and explores a potential computational approach to achieving it.

%% file: 03-related.tex
\section{Related work}
\label{sec:related}

\vspace{.5em}
\noindent\textbf{Exposing query identification. }
\label{sec:related-exposure}
The exposing query identification problem was defined by in the context of content creator privacy~\citet{biega2017learning}.
The authors proposed an EQI approach that identifies unigram and bigram exposing queries in the specific case of a ranker that uses unigram language models with Dirichlet smoothing. The solution generates synthetic candidate queries from the corpus and then prunes this queryset using a ranker-specific heuristic. The output exposing query set is exact, although computed over a limited set (unigram and bigram) of not always realistic synthetic queries.
Instead, formulating EQI as a retrieval task over a corpus of queries allows us to:
 \begin{enumerate*}[label=(\roman*)]
 \item model content exposure more adequately by incorporating user browsing models; and
 \item expand the space of solutions to include query retrieval methods that can handle arbitrary-length exposing queries.
 \end{enumerate*}

EQI is closely related to the concept of \emph{retrievability} developed by~\citet{azzopardi2008retrievability}. Retrievability is a dot product score assigned to a document quantifying whether (1) a document is exposed within a rank a user is patient enough to inspect and (2) how likely a query is to be issued to the search engine. Our goal is to discover specific queries which expose a document to a searcher and could thus be used to estimate component (1) of retrievability.
Further related concepts include \emph{keyqueries}~\cite{gollub2013keywords}, which are the minimal queries retrieving a document in a given document retrieval system. By definition, keyqueries are thus a subset of all exposing queries.

\vspace{.5em}
\noindent\textbf{Search transparency and explainability. }
EQI is a form of search \emph{transparency}: the goal is to help users understand comprehensively what outcomes the search system yields for them. Transparency is closely related to interpretability and explainability, where the goal is to understand how and why a ranking algorithm returns certain results.
Explainability approaches include generation of post-hoc explanations for black-box models, both local (explaining a single ranking result)~\cite{verma2019lirme,singh2019exs} and global (explaining a ranking model)~\cite{singh2018posthoc}. 
Model-oriented techniques include modifying models to be more interpretable and transparent~\cite{zhuang2020interpretable} or explaining through visualizations~\cite{chios2021helping,choi2020interpreting}, while user-oriented techniques include inferring the intent behind a search query that a model has assumed~\cite{singh2018interpreting} or tailoring explanations to user mental models of search~\cite{thomas2019investigating}, evaluating explanations through the lens of trust~\cite{polley2021towards}.

\vspace{.5em}
\noindent\textbf{Reversing the role of queries and documents. }
\label{sec:related-related}
Our approach to EQI is inspired by several IR problems in which the role of documents and queries is reversed. \citet{yang2009query} employ a query-by-document technique issuing the most informative phrases from a document as queries to identify related documents. In the context of EQI, the setup has a few differences: We retrieve exposing queries rather than other related documents, and are interested in an extensive coverage of exposing queries thus using the whole document as a query rather than just the most informative phrases. \citet{pickens2010reverted} index query result sets, where each retrievable item is a query and each query is described by a list of pseudo-terms that correspond to document IDs that the query retrieves using a chosen retrieval system.
This approach, referred to as \emph{reverted indexing}, allows for query-by-multiple-documents. In our context, where we are interested in retrieving queries that expose a single document, reverted indexing is akin to the exact brute-force method.

\citet{santos2013learning} develop a framework for suggesting queries which might retrieve more relevant results for a given query. \citet{nogueira2019document} develop a generative model which, for a given document, produces queries that could expand the document for more effective retrieval. Both approaches produce a set of queries to which a given document may be \emph{relevant}, but not those that would expose the document when issued specifically to a given document retrieval system.
We expect that the set of queries generated by these approaches to intersect with the ideal set of exposing queries, but as the behavior of the particular document retrieval system diverges from the labeled data, the intersection will be smaller; and the effectiveness may also vary between labeled-relevant, labeled-nonrelevant, and unlabeled documents.
It is important in these cases for EQI to be effective for documents that are actually exposed, irrespective of their relevance.

%% file: 04-task.tex
%\vspace{-5px}
\section{Exposing query identification}
\label{sec:task}

We are given a \emph{document retrieval} system producing a ranked list of documents retrieved from a collection $\mathcal{D}$ in response to a query.
We define EQI as a complementary retrieval task where, given a document $d$, the system is responsible for retrieving a ranked list of queries from a query log $\mathcal{Q}$, ranked according to their probability of exposing the document in the document retrieval system.
Thus, EQI technically means reversing document retrieval.
We use subscripts $\textbf{\fbox{}}_{\;q \shortto d}$ and $\textbf{\fbox{}}_{\;d \shortto q}$ to denote variables corresponding to the document retrieval and the EQI task, respectively.
Table~\ref{tbl:notation} presents the notations.

\begin{table}
    \small
    \centering
    \caption{List of notations.}
    \label{tbl:notation}
    \begin{tabular}{ll}
    \toprule
        \textbf{Notation} & \textbf{Description} \\
    \midrule
    $\mathcal{D}$ & A collection of documents \\
    $\mathcal{Q}$ & A corpus of queries\\
    $d$ & An individual document \\
    $q$ & An individual query \\
    \midrule
    \multicolumn{2}{l}{\textbf{Document retrieval}} \\
    $\sigma_q$ & A ranked list of documents retrieved in response to $q$ \\
    $\sigma_q^*$ & The ideal set of documents for  $q$\\
    $n_{q \shortto d}$ & Number of documents retrieved per query\\
    $\mu_{q \shortto d}$ & A user browsing model for inspecting $\sigma_q$ \\
    $\gamma_{q \shortto d}$ & Persistence parameter for inspecting $\sigma_q$\\
    $\rho(d,\sigma_q)$ & 0-based rank of $d$ in $\sigma_q$\\
    \midrule
    \multicolumn{2}{l}{\textbf{Exposing query retrieval}} \\
    $\psi_d$ & A ranked list of queries retrieved in response to $d$ \\
    $\psi_d^*$ & The ideal ranked list of queries for $d$ \\
    $n_{d \shortto q}$ & Number of queries retrieved per document\\
    $\mu_{d \shortto q}$ & A user browsing model for inspecting $\psi_d$ \\
    $\gamma_{d \shortto q}$ & Persistence parameter for inspecting $\psi_d$\\
    $\rho(q,\psi_d)$ & 0-based rank of $q$ in $\psi_d$\\
    \bottomrule
    \end{tabular}
\end{table}

\subsection{Deriving an Evaluation Metric}

Retrieval metrics often take the form of a cumulative quality score summed over all items in the top-ranking positions weighted by the probability that a searcher inspects the document at a given rank:
%
%\begin{align}
$
    \mathcal{M}(\sigma) = \sum_{d \in \sigma}{\mu(d, \sigma)\cdot g_d}. 
$
%\end{align}
%
Here, $\sigma$ is the result list, the quality score $g_d$ of a document $d$ is typically a function of its relevance to the query and $\mu(d, \sigma)$ is the probability of a user inspecting $d$ under the user browsing model $\mu$.
Metrics---such as, NDCG~\citep{jarvelin2002cumulated}---further normalize this value by the ideal value of the metric: 
%
%\begin{align}
$
    \text{Norm-}\mathcal{M}(\sigma) = \frac{\sum_{d_i \in \sigma}{\mu(d_i, \sigma)\cdot g_{d_i}}}{\sum_{d_j \in \sigma^*}{\mu(d_j, \sigma^*)\cdot g_{d_j}}}
    \label{eqn:norm-ir-metric}
$
%\end{align}
%
Where $\sigma^*$ is the ideal ranked list of results.
Similarly, in the exposing query retrieval scenario, we want to measure the quality of the ranked list of retrieved queries, which we refer to as the exposure list, in response to a document.
Towards that goal, we propose the Ranked Exposure List Quality (RELQ), which assumes a similar form as $\text{Norm-}\mathcal{M}$:
% \vspace{-3px}
\begin{align}
    \text{RELQ}_{\mu_{d \shortto q}}(\psi_d) &= \frac{\sum_{q_i \in \psi_d}{\mu_{d \shortto q}(q_i, \psi_d)\cdot g_{q_i}}}{\sum_{q_j \in \psi^*_d}{\mu_{d \shortto q}(q_j, \psi_d^*)\cdot g_{q_j}}}
    \label{eqn:resq-prenorm}
\end{align}
Where $\psi_d$ and $\psi_d^*$ are the retrieved and ideal exposure list for $d$. \newline

The \emph{key observation} now is that in case of EQI, the quality value $g_q$ corresponds to the probability that $q$ exposes $d$ in the original retrieval system---which is given by $\mu_{q \shortto d}$:
% \vspace{-3px}
\begin{align}
    \text{RELQ}_{\mu_{d \shortto q}, \mu_{q \shortto d}}(\psi_d) &= \frac{\sum_{q_i \in \psi_d}{\mu_{d \shortto q}(q_i, \psi_d)\cdot \mu_{q \shortto d}(d, \sigma_{q_i})}}{\sum_{q_j \in \psi^*_d}{\mu_{d \shortto q}(q_j, \psi_d^*)\cdot \mu_{q \shortto d}(d, \sigma_{q_j})}}
    \label{eqn:resq}
\end{align}

This derivation leads us to a metric that jointly accounts for the behavior of two distinct users: 
\begin{enumerate*}
\item the searcher using the document retrieval system and to whom the content is exposed, and 
\item the user of the EQI system.
Intuitively, the proposed metric has a higher value when the EQI user is exposed to queries that in turn prominently expose the target document in their corresponding rank lists.
\end{enumerate*}\footnote{
Similar two-sided metrics have previously been used in the task of query suggestion: the success of a query prediction system depends on the quality of the query ranking but also on the quality of the document ranking each query retrieves~\cite{santos2013learning}. EQI requires a similar multiplicative metric; unlike the query prediction metric, however, an exposure set quality metric needs to:
\begin{enumerate*}[label=(\roman*)]
\item quantify whether a query \emph{exposes} a document rather than whether the document is relevant; and
\item incorporate two different models of user behavior (EQI and document retrieval system users).
\end{enumerate*}}

\vspace{5px}
\noindent\textbf{Metric instantiations. }To compute the metric, we can plug in standard user behavior models for $\mu_{d \shortto q}$ and $\mu_{q \shortto d}$ that make different assumptions about how users interact with retrieved results under the document retrieval and the exposing query retrieval settings.
For example, if we assume that both $\mu_{d \shortto q}$ and $\mu_{q \shortto d}$ are based on the same user model as the Rank-Biased Precision (RBP) metric~\citep{Moffat:2008:RBP} with persistence parameters $\gamma_{d \shortto q}\in(0,1]$ and $\gamma_{d \shortto q}\in(0,1]$, we get the following variant: %of the RELQ metric:
% \vspace{-5px}
\begin{align}
    \text{RELQ}_\text{RBP, RBP}(\psi_d) &= \frac{\sum_{q_i \in \psi_d}{\gamma_{d \shortto q}^{\rho(q_i, \psi_d)}\cdot \gamma_{q \shortto d}^{\rho(d, \sigma_{q_i})}}}{\sum_{q_j \in \psi^*_d}{\gamma_{d \shortto q}^{\rho(q_j, \psi_d^*)}\cdot \gamma_{q \shortto d}^{\rho(d, \sigma_{q_j})}}}
    \label{eqn:resq-rbprbp}
\end{align}

Where, $\rho(x, \Omega)$ is the $0$-based rank of $x$ in the ranked list $\Omega$.
% , and we define $\rho(x, \Omega)=\infty$ if $x \notin \Omega$.

% Alternatively, we can assume that the user exhaustively inspects all results in the exposure query retrieval setting, while still adopting the RBP model for document retrieval.
% This is exactly the special case where $\gamma_{d \shortto q}=1$.
% \vspace{-5px}
% \begin{align}
%     \text{RELQ}_\text{EXH, RBP}(\psi_d) &= \frac{\sum_{q_i \in \psi_d}{\gamma_{q \shortto d}^{\rho(d, \sigma_{q_i})}}}{\sum_{q_j \in \psi^*_d}{\gamma_{q \shortto d}^{\rho(d, \sigma_{q_j})}}}
%     \label{eqn:resq-exhrbp}
% \end{align}

% If we assume that the user inspects retrieved results exhaustively under both retrieval settings, the metric formulation simplifies to:
% \vspace{-2px}
% \begin{align}
%     \text{RELQ}_\text{EXH, EXH}(\psi_d) &= \frac{\sum_{q_i \in \psi_d}{\mathbbm{1}\{d \in \sigma_{q_i}\}}}{\sum_{q_j \in \psi^*_d}{\mathbbm{1}\{d \in \sigma_{q_j}\}}}
%     \label{eqn:resq-exhexh} 
% \end{align}

% Where, $\mathbbm{1}\{.\}$ is the indicator function.

Instead of RBP, we can plug in different combinations of user models, including NDCG or exhaustive (where a user inspects all ranking results). For example, if we
%
%We can also 
assume an exhaustive model for EQI, while adopting NDCG for the document retrieval:
% \vspace{-3px}
\begin{align}
    \text{RELQ}_\text{EXH, NDCG}(\psi_d) &= \frac{\sum_{q_i \in \psi_d} \vmathbb{1}\{d \in \sigma_{q_i}\} \cdot \frac{\rho(d, \sigma_{q_i})}{\log_2(i+1)}}
    {\sum_{q_j \in \psi^*_d}{\vmathbb{1}\{d \in \sigma_{q_j}\}} \cdot \frac{\rho(d, \sigma_{q_j})}{\log_2(j+1)}}
    \label{eqn:resq-exhrbp}
\end{align}
Where, $\vmathbb{1}\{.\}$ is the indicator function.

\subsection{Practical considerations}
\subsubsection{Prototypical users}
\label{sec:typical_users}

$RELQ_\text{RBP, RBP}$ is parametrized by two user patience parameters: $\gamma_{q \shortto d}$ and $ \gamma_{d \shortto q}$. Their values should be chosen to best reflect the behavior of users in a given underlying task. Absent behavioral data, we can instead represent certain prototypical users. Patient document retrieval users might be curious or malicious (in case of positive or negative query contexts, respectively). Patient EQI system users might be worried about their privacy or interested in monetization of their content. 
% In this paper, we provide experimental analyses for various pairs of users, including exhaustive explorers ($\gamma=1.0$), as well as high-patience ($\gamma=0.9$), and average-patience ($\gamma=0.5$) users.

\subsubsection{Query collections}
EQI assumes the existence of an exhaustive query collection. There are a number of approaches to creating such a collection. First, we can use an existing log from the document retrieval system that contains all the queries ever issued by searchers. The advantage of this approach is that the queries will be realistic. However, the model will not be able to capture instances of problematic exposure~\cite{biega2016r,biega2017learning,fbLeak} by unseen queries. Synthetically generating a collection of queries is an alternative. To limit the collection size when this approach is adopted, queries can be truncated at a certain length and further pruned according to various frequency heuristics ~\cite{azzopardi2008retrievability,biega2017learning,callan2001query}. The approach might allow the system to detect problematic exposure before it occurs, but it will prevent the system from surfacing exposure to many viable queries. A practitioner should choose an approach depending on whether it is important to report the common or worst-case exposure contexts in a given application.

%% file: 06-ance.tex
\section{EQI for embedding-based search systems}
\label{sec:ance}

The first family of retrieval systems that we consider in this work are embedding-based search systems, also referred to as dual-encoder systems, that learn independent vector representations of query and document such that their relevance can be estimated by applying a similarity function over the corresponding vectors.
Let $f_Q : \mathcal{Q} \to \mathbb{R}^n$ and $f_D : \mathcal{D} \to \mathbb{R}^n$ be the query and document encoders, respectively, and $\otimes: \mathbb{R}^n \times \mathbb{R}^n \to \mathbb{R}$ be the vector similarity operator.
The score $s_{q,d}$ for a query-document pair is then given by,
\begin{align}
    s_{q,d} &= f_Q(q) \otimes f_D(d)
\end{align}

Ideally, $s_{q,d_+} > s_{q,d_-}$, if $d_+$ is more relevant to the query $q$ than $d_-$.
While $\otimes$ is typically a simple function, such as dot-product or cosine similarity, the encoding functions $f_Q$ and $f_D$ are often learned by minimizing an objective of the following form:
\begin{align}
    \mathcal{L}_{q \shortto d} &= \mathbb{E}_{q, d_+, d_- \sim \theta} [\ell_{q \shortto d}(s_{q,d_+} - s_{q,d_-})]
    \label{eqn:loss-fwd}
\end{align}

Commonly used functions for the instance loss $\ell_{q \shortto d}$ include hinge \citep{herbrich2000large} or RanknNet \citep{burges2005learning} loss.
For example, in case of RankNet:
\begin{align}
    \ell_{q \shortto d}(\Delta) &= \text{log}(1 + e^{-\Delta})
    \label{eqn:inst-loss-fwd}
\end{align}

\textbf{Metric learning for EQI. }
For EQI, we want to learn a similarity metric such that a query $q^+$ should be more similar to document $d$ than $q^-$, if $d$ has a higher probability of exposure to the user on the search result page $\sigma_{q^+}$ compared to $\sigma_{q^-}$.
Similar to the embedding-based model for document retrieval, we now define two new encoders $h_Q : \mathcal{Q} \to \mathbb{R}^n$ and $h_D : \mathcal{D} \to \mathbb{R}^n$ for query and document, respectively, and write our new optimization objective as follows:
\begin{align}
    \mathcal{L}_{d \shortto q} &= \mathbb{E}_{d, q_+, q_- \sim \theta} [\ell_{d \shortto q}(u_{d, q_+} - u_{d, q_-})]
    \label{eqn:loss-rev}
\end{align}

Where, $ u_{d, q} = h_Q(q) \otimes h_D(d)$.
If we assume that exposure of a document depends only on the rank position at which it is displayed on a result page, then $q_+$ and $q_-$ should be sampled such that $\rho(d,\sigma_{q_+}) < \rho(d,\sigma_{q_-})$---\ie, $d$ is displayed at a higher rank position for $q_+$ than $q_-$.
Let, $X_{q \shortto d} = \{f_Q(q) | q \in \mathcal{Q}\} \bigcup \{f_D(d) | d \in \mathcal{D}\}$ and $X_{d \shortto q} = \{h_Q(q) | q \in \mathcal{Q}\} \bigcup \{h_D(d) | d \in \mathcal{D}\}$.
Then, $(X_{q \shortto d}, \otimes)$ and $(X_{d \shortto q}, \otimes)$ define two distinct metric spaces corresponding to the embedding-based document retrieval system and exposing query retrieval system, respectively.
To retrieve exposing queries for a document $d$, we need to perform a reverse nearest neighbor (rNN) search in the $(X_{q \shortto d}, \otimes)$ metric space.
However, rNN solutions are inefficient for high-dimensional and dynamic models.
We instead propose to learn a new metric space $(X_{d \shortto q}, \otimes)$ where a NN search would approximate the rNN search in the $(X_{q \shortto d}, \otimes)$ space.

\subsection{Training data generation}
\label{sec:ance-training-data}
Generating training data is a key challenge in this learning task.
To train the model, we need a dataset of documents and their exposing queries.
We thus need to devise a training data generation strategy which ensures that: (1) we only issue a limited number of queries to the document retrieval model to keep the approach computationally efficient, (2) we have at least one exposing query for each training document, and (3) the training data represents the distribution of queries around each training document. The pseudocode for the proposed procedure is outlined in Algorithm \ref{alg:train-data}.
Training data is ensured to be representative and high-quality under strict efficiency constraints as we only conduct document retrieval for the training queries (line~\ref{algo:line:5}). By selecting training documents from the candidate set $\mathcal{D}_{\text{candidate}}$ (line~\ref{algo:line:9}), we guarantee that each training document has at least one exposing query that retrieves it in top ranks. We then do a KNN search from $Q_{\text{train}}$ of the queries closest to each training document in the document retrieval embedding space (line~\ref{algo:line:12}). This gives us a sample distribution of queries around each document and the positive and negative document-query pairs will match the current errors in the document retrieval embedding space (where similarity between document and query does not necessarily reflect the exposure).

Note that from $\Psi_{\text{train}}$, as constructed in Algorithm \ref{alg:train-data}, we can sample two types of training instances of form $<d, q^+, q^->$, corresponding to:
\begin{enumerate*}[label=Case \arabic*:,leftmargin=4\parindent]
    \item $\rho(d,\sigma_{q^+}) < \rho(d,\sigma_{q^-}) < n_{q \shortto d}$, and 
    \item $\rho(d,\sigma_{q^+}) < n_{q \shortto d} \le \rho(d,\sigma_{q^-})$.
\end{enumerate*}
During training, we randomly sample training instances for Case 1 and Case 2 with probability $\alpha$ and $1 - \alpha$, respectively. Intuitively, Case 1 helps the model better rank the exposing queries already retrieved by the document retrieval embedding space, while Case 2 helps the model learn to retrieve exposing queries that are far away from the document in the starting document embedding space.

\begin{algorithm}[t]
\caption{Training data generation}
\label{alg:train-data}
\begin{algorithmic}[1]
\STATE Randomly sample $\mathcal{Q}_{\text{train}}$ from $\mathcal{Q}$
\STATE Initialize $\mathcal{D}_{\text{candidate}} = \emptyset$ 
\STATE Initialize cache results $\mathcal{C}=\{\}$
\FORALL{$q_\text{train} \in \mathcal{Q}_{\text{train}}$}
\STATE Retrieve top $n_{q \shortto d}$ ranked list of documents $\sigma_{q_{\text{train}}}$ from $\mathcal{D}$ \label{algo:line:5}
\STATE Cache retrieval results $\mathcal{C}[q_{\text{train}}]=\sigma_{q_{\text{train}}}$
\STATE $\mathcal{D}_{\text{candidate}} \leftarrow \mathcal{D}_{\text{candidate}} \bigcup \; n_{q \shortto d} \; \text{documents in} \; \sigma_{q_{\text{train}}}$
\ENDFOR
\STATE Randomly sample $\mathcal{D}_{\text{train}}$ from $\mathcal{D}_{\text{candidate}}$ \label{algo:line:9}
\STATE Initialize training dataset $\Psi_{\text{train}} = \{\}$ 
\FORALL{$d_\text{train} \in \mathcal{D}_{\text{train}}$}
\STATE Retrieve top $n_{d \shortto q}$ ranked list of queries $\psi_{d_{\text{train}}}$ from ${\mathcal{Q}_{\text{train}}}$ \label{algo:line:12}
\STATE Initialize $\beta_{d_{\text{train}}}=\{\}$
\FORALL{$q \in \psi_{d_{\text{train}}}$ }
\IF{$q \in \mathcal{C}[q]$}
\STATE Get document rank $\rho(d_{\text{train}}, \sigma_q)$ from $\mathcal{C}[q]=\sigma_q$
\ELSE{}
\STATE Label $\rho(d_{\text{train}}, \sigma_q)$ as $\infty$
\ENDIF
\STATE $\;\;\;\; \beta_{d_{\text{train}}}[q]=\rho(d_{\text{train}}, \sigma_q)$ 
\ENDFOR
\STATE Add to the training dataset $\Psi_{\text{train}} [d_{\text{train}}]=\beta_{d_{\text{train}}}$
\ENDFOR
\RETURN $\Psi_{\text{train}}$
\end{algorithmic}
\end{algorithm}

\subsection{Model architecture}
\label{sec:ance-network}
We adopt the publicly available\footnote{\href{https://github.com/microsoft/ANCE}{https://github.com/microsoft/ANCE}}
BERT-based dual-encoder architecture by \citet{xiong2020approximate} as our dual-encoder search system.
The model is trained in an active metric learning setting~\citep{kumaran2018active, yang2012bayesian, ebert2012active}, optimizing towards the so-called approximate nearest neighbor negative contrastive estimation (ANCE) objective.
As far as we are aware, the model trained on the ANCE loss has demonstrated a state-of-the-art performance by a dual-encoder system. We therefore adopt this model as our base architecture for both the document and the exposing query retrieval stages.
Unless specifically mentioned, we use the same hyperparameters as in the original paper.

We anticipate that both models---for the document retrieval and the exposing query retrieval tasks---need to learn a fundamental notion of topical relevance between queries and passages that are largely similar in both cases.
Therefore, we pre-initialize the exposing query retrieval model with the learned parameters of the document retrieval model before training. 
Furthermore, we add a few additional layers on top of the query and document encoders. However, efficiency constraints limit the amount of data we can use for training the extra layers. To avoid overfitting, we design the network architecture to only slightly transform the original embedding space. In particular, we explore following two settings:

\textit{Append:} 
Let $\vec{x}$ be the vector output of the base encoder, which for our base model has a size of $768$.
Then, under the Append settings, the final embedding output is $\vec{y} = \big(\vec{x}, \text{FF}(\vec{x})\big)$, where $(,)$ denotes a concatenation of two vectors, and FF is a feedforward network.
Specifically, in our experiments we employ a four-layer feedforward neural network, with hidden layer sizes of $64$ and the last output dimension as $32$.
Consequently, after concatenation the final embedding vector dimensions is $800$.
For each hidden layer of the sub-network, we apply a ReLU nonlinearity, a dropout layer, and a layer normalization.

\textit{Residual:}
Using a similar notation, the output of the residual layer~\cite{he2016deep} can be written as $\vec{y} = \vec{x} + \text{FF}(\vec{x})$.
The feedforward network FF in this case comprises a single hidden layer of size $384$ and an output vector size of $768$.
We add ReLU nonlinearity, dropout, and layer normalization similar to the Append setting.

\subsection{Training and retrieval}
\label{sec:ance-train}

We train all models, including baselines, for 1K iterations, where each iteration consists of $100$ batches with $1000$ samples per batch. We use dot product as the similarity measurement as in the ANCE model. We use a pairwise optimization objective in the form of the cross entropy loss.
We set $\alpha$, as introduced in Section \ref{sec:ance}, to 0.5 and the dropout rate to $0.1$.
We use Adam optimizer with a fixed learning rate of 1e-4---and set the coefficients beta1 to 0.9, beta2 to 0.999, and eps to 1e-08. Unless otherwise specified, all results reported in Section \ref{sec:result} are obtained from models trained on 50K queries and 200K passages.

For experimentation agility under limited GPU budget, we freeze the base encoders and only optimize the parameters of the FF layers on top---under both Append and Residual settings---when training for the exposing query retrieval task.
We expect that additional improvements may be possible from fine-tuning the encoders end-to-end but we do not explore that direction in our current work.

We use the Faiss IndexFlatIP Index \cite{johnson2019billion}, which guarantees efficient computation of exact nearest neighbors, for nearest neighbor search in the context of both our baseline and proposed models, as well as for the ground truth results generation.

%% file: 05-bm25.tex
\section{EQI for traditional search systems}
\label{sec:bm25}

%In traditional TF-IDF-based retrieval methods~\citep{salton1988term}, such as BM25~\citep{robertson2009probabilistic}, the query and document are represented by weighted term vectors $x_q \in \mathbb{R}^{|T|}$ and $x_d \in \mathbb{R}^{|T|}$, where $T$ is the set of all unique terms in the collection $\mathcal{D}$.
%The query-document relevance score $s_{q,d}$ is estimated using simple similarity metric $\otimes: \mathbb{R}^{|T|} \times \mathbb{R}^{|T|} \to \mathbb{R}$, such as dot-product.
%
%\begin{align}
%$
%    s_{q,d} = x_q \otimes x_d 
%            = \sum_{t}^{T}{x_{qt} \cdot x_{dt}}.
%$
%\end{align}
%
%
We also study EQI in the context of BM25~\citep{robertson2009probabilistic}, an important traditional retrieval model.
BM25 estimates the query-document relevance score as a function of how often the term occurs in the text (term frequency or TF) and some measure of how discriminative the term is---\eg, inverse-document frequency (IDF)~\citep{robertson2004understanding}.
\begin{align}
    s_{q,d} = \sum_{t \in q}{IDF(t) \cdot \frac{TF(t, d) \cdot (k_1+1)}{TF(t, d) + k_1 \cdot (1 - b + b \cdot \frac{|d|}{\text{AvgDL}})}}
\end{align}
Where, $k_1$ and $b$ are parameters of the model that can be tuned and AvgDL is the average document length in the collection.
% The term-document scores are typically precomputed and encoded as posting lists in specialized data structures, such as inverted index.
% At query time, only the term-query weights are computed and subsequently the top-$k$ documents are retrieved by searching over the positing lists~\citep{zobel2006inverted}.
A simple approach to retrieving exposing queries in the context of BM25 may therefore involve simply exchanging the role of query and documents under this setting.
Instead of precomputing term-document scores and indexing the document collection, we can instead precompute the term-query scores and index the query log.
Given the weighted term vector corresponding to a document, we can retrieve queries from the prepared index.
Real life IR systems often make practical assumptions about the queries being typically short and documents being longer in their design to achieve more efficient retrieval.
While we may need to work around some of these system constraints, the general framework for reversing a BM25-based system seems straightforward.
The term weighting functions can be further tuned, if necessary, for the exposing query retrieval setting independently of the scoring function employed for document retrieval.

\subsection{Indexing and retrieval}
\label{sec:bm25-implementation}
In our work, we adopt the Anserini toolkit~\cite{yang2017anserini} to index the query log and issue documents as queries.
We tune the model parameters, observing that the impact is limited under the reversed setting as the term-saturation is applied on the short indexed queries. We foresee future work that applies the term-saturation on the input to the retrieval system instead, but that consequently requires the retrieval system to support explicit term-weighting.

%We do not tune any of the model parameters but foresee that future work can improve the quality of exposing query retrieval by tuning such parameters and explicitly specifying term weights in the query.
%To explore the performance of reversing a traditional BM25 document retrieval model, we take the simple approach of exchanging query and documents. We adopted the Anserini toolkit~\cite{yang2017anserini} to perform both the document retrieval and exposing query retrieval task. The document retrieval builds an inverted index on the documents and used queries to retrieve a ranked list. And for the exposing query retrieval task, we build an inverted index using the queries, and apply documents as the query to do the retrieval.

%% file: 07-experiment.tex
\section{Experiments}
\label{sec:experiment}

\subsection{Data}
\label{sec:method-data}
For EQI experimentation, we need a large-scale dataset accompanied by a large query collection. We thus use the MS MARCO passage retrieval benchmark~\citep{bajaj2016ms}, extensively used to compare neural methods for retrieval, including in the TREC Deep Learning track~\citep{craswell2020overview}.
The MS MARCO dataset contains  $8,841,823$ passages and $532,761$ relevant query-passage pairs for training, as labeled by human assessors.

\subsection{Methods}
\label{sec:experiment-baseline}

\paragraph{Brute-Force (Exact EQI) Baseline} 
An obvious baseline for our work corresponds to the brute-force approach of running every query in our log through the document retrieval system and recording all the retrieved results.
Then for a given passage, we can iterate over all the document retrieval results to produce an exact solution to the exposing query retrieval problem.
While the brute force method achieves perfect recall of all exposing queries, it also incurs both a heavy computation cost for the document retrieval step.

% and a significant storage cost for caching of the results for future lookup.
% Furthermore, there is limited reusability under this approach---if we replace the one query log for another, it would require significant re-computation even if the new query log is distributionally similar to the old one.

% \paragraph{Approximate nearest neighbor (ANN) search in document retrieval embedding space:}
% As mentioned in Section~\ref{sec:related-exposure} and Section~\ref{sec:model}, the exposing query identification challenge can be solved exactly by performing a reverse nearest neighbor search in the embedding space learned by the document retrieval system, albeit at the high computational cost. In this paper, we are interested in lightweight and approximate solutions.
% In the context of embedding-based search systems, the core of our approach is in transforming a document retrieval embedding space into an exposing query embedding space. As argued before, such a transformation is needed since the nearest neighbor search and the reverse nearest neighbor search problems are generally not symmetrical. Still, we expect that even in the document retrieval embedding space, exposing query and document pairs would be relatively close to each other, and a nearest neighbor search around a passage in this space would identify some of its exposing queries. We employ this approach as a baseline, which we refer to as the ANN-Orig baseline in the remainder of this paper.

\paragraph{Original Retrieval Space Methods.}
For traditional search systems, an intuitive baseline model can index the query log as a document collection and use documents as queries. We refer to this method as BM25-reverse. We also tune the term weighting parameters $k_1$ and $b$, which we refer to as BM25-tuned. In the context of dual-encoder search systems, 
% the core of our approach is in transforming a document retrieval embedding space into an exposing query embedding space. As argued before, such a transformation is needed since the nearest neighbor search and the reverse nearest neighbor search problems are generally not equivalent. 
we expect that, even in the document retrieval embedding space, exposing query and document pairs would be relatively close to each other, and a nearest neighbor search around a passage in this space would identify some of its exposing queries. We employ this approach as a baseline, which we refer to as the ANCE-reverse in the remainder of this paper.

\paragraph{Metric learning. } We also implement the metric learning approaches described in Sec.~\ref{sec:ance}, referring to the two architectures as ANCE-append and ANCE-residual.

\subsection{Ground truth end experimental protocol} 
\label{sec:method-data}
 In our experiments, we set $n_{q\shortto d}=100$ and $n_{d\shortto q}=100$. 
 To obtain the ground truth exposing queries we obtained a ranked list $\sigma^*_q$ of the top 100 passages for each query $q$ in the collection.
 %To obtain the ground truth exposing queries, for each query $q$ in the query collection, we obtained a ranked list $\sigma^*_q$ of the top 100 nearest documents in the document retrieval embedding space, and then brute-force reversed the results.
%  We pair $\sigma_q$ with $q$, so that $\psi^*_d$ can be easily found by reversing the key as passage $d$. 
 To evaluate the EQI models, for each model and each passage $d$ in the test set of size $20,000$, we retrieve the top 100 queries as the exposing ranked list $\psi_d$.
%The experiments of this paper are conducted on the corpus of passage retrieval track of MS MARCO dataset\cite{bajaj2016ms}. The dataset consists of 8,841,823 passages and 1,010,916 queries. In our experiments, we selected whole passage corpus and the training query set with 532,761 queries. As the dual-encoder dropped some of the queries formed by rare terms, the final number of queries used in the following experiment on exposing query retrieval is 502,939.   \asia{I'm not sure I get this previous sentence---was something left out from the dataset?}\jl{Firstly, we started from the raw text of (queries.train.tsv) training queries of size 808,731 in our experiments, and didn't include the dev and test set queries. Then, among the 808,731 train queries, there are only 532,761 of them that are included in the qrels(qrels.train.tsv). As some of the queries is fully made of rare terms that are not embedded, the final number of valid queries that can be embedded and used in our experiments is 502,939. I'm not sure about if we want to talk about this process in detail, as it is more about a forward ranker implementation thing.}\asia{We need to include all the details necessary for the readers to replicate our experiments.}

%% file: 08-result.tex
\section{Results}
\label{sec:result}

\subsection{Overall performance}
\label{sec:model_performance}

\begin{table}
  \small
  \begin{center}
  \caption{Model performance on the EQI task benchmarked using the MS MARCO dataset.
  We report the RELQ metric corresponding to four different user model combinations.
  }
 \label{tab:main_results}
    \begin{subtable}{\columnwidth}
    \begin{tabular}{l c c c c}
    \toprule
       & \multicolumn{3}{c}{$\text{RELQ}_\text{RBP,RBP}$ with ($\gamma_{q \shortto d}$, $\gamma_{d \shortto q}$)}& $\text{RELQ}_\text{EXH,NDCG}$\\ %\cline{2-5}
      Model & $(0.5, 0.5)$ & $(0.5,0.9)$ & $(1,1)$ \\ 
      \midrule
      BM25-reverse & $0.441$ & $0.624$ & $0.840$ & $0.645$ \\
      BM25-tuned & $0.442$ & $0.626$ & $0.845$ & $0.648$ \\
      Brute force & $1.000$ & $1.000$ & $1.000$ & $1.000$\\
    \bottomrule
    \end{tabular}
    \caption{EQI for BM25}
    \end{subtable} 
    \begin{subtable}{\columnwidth}
    \vspace{4px}
    \begin{tabular}{l c c c c}
    \toprule
       & \multicolumn{3}{c}{$\text{RELQ}_\text{RBP,RBP}$ with ($\gamma_{q \shortto d}$, $\gamma_{d \shortto q}$)}& $\text{RELQ}_\text{EXH,NDCG}$\\ %\cline{2-5}
      Model & $(0.5, 0.5)$ & $(0.5,0.9)$ & $(1,1)$ \\
      \midrule
      %\multicolumn{5}{l}{\textbf{Baseline}} \\
      ANCE-reverse & $0.493$ & $0.685$ & $0.887$ & $0.698$ \\
      %\midrule
      %\multicolumn{5}{l}{\textbf{Our models}} \\
      ANCE-append & $0.624$ & $0.825$ & $0.982$ & $0.837$ \\
      ANCE-residual & $0.633$ & $0.834$ & $0.984$ & $0.845$ \\
      Brute force & $1.000$ & $1.000$ & $1.000$ & $1.000$ \\
    \bottomrule
    \end{tabular}
    \caption{EQI for ANCE}
    \end{subtable}
  \end{center}
\end{table}

Table~\ref{tab:main_results} summarizes the EQI performance results on the MS MARCO benchmark, using different parametrizations of the RELQ metric.
The brute force baseline, by definition finding all exposing queries, achieves a perfect value of the RELQ metric.
In the context of traditional search models, BM25-tuned consistently shows a slightly better performance than BM25-reverse according to all metrics. In the context of dual-encoder models, both the proposed metric learning approaches (ANCE-append and ANCE-residual) significantly outperform the ANCE-reverse under different parametrizations of RELQ.
The improvements of both models against ANCE-reverse are statistically significant according to a one-tailed paired t-test ($p < 0.01$).
Of the two models, Residual performs better than Append, and the difference in their performance are again statistically significant based on  one-tailed paired t-test ($p < 0.01$).
Therefore, we only use the Residual model in the remaining analyses.

\begin{figure}
    \centering
    \includegraphics[width=0.9\linewidth]{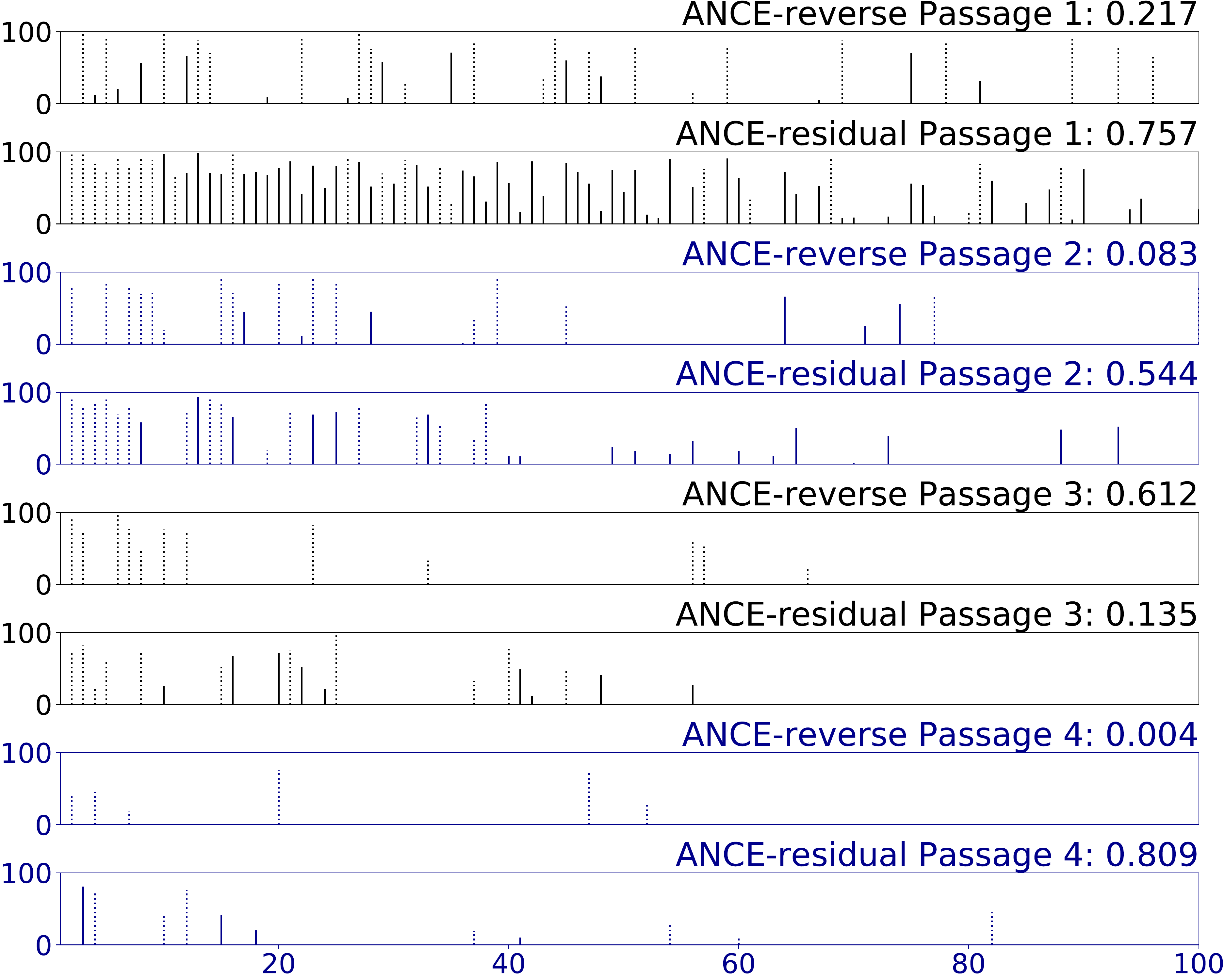}
    \caption{Visualization of retrieved exposing query lists. The figure presents retrieved ranked lists of exposing queries for four test passages. Dotted lines represent the exposing queries retrieved by both models. The number reported is the $RELQ_{RBP,RBP}$ with $\gamma_{q\shortto d}=0.5, \gamma_{d\shortto q}=0.9$. The x-axis represents the ranking position in the exposing query retrieval model $\rho(q,\psi_d)+1$ and the y-axis represents the rank at which the query retrieves the test passage in the document retrieval system $n_{q\shortto d}-\rho(d,\sigma_q)$ (the higher, the more exposing).} 
    \label{fig:example}
\end{figure}

\paragraph{Illustration. }Beyond the mean metric values, we illustrate the behavior of ANCE-reverse and ANCE-residual on four selected test passages in Figure \ref{fig:example}.
These passages were selected to reflect different performance levels.
For each passage, we compare the ranked exposing query lists retrieved by the ANCE-Reverse and the Residual models. Recall that $n_{d\shortto q}=100$.
Vertical bars denote exposing queries, with the dotted vertical bars showing queries retrieved by both models.
The x-axis represents the ranking position of the query in the EQI model, while the y-axis represents the ranking position at which the document retrieval system retrieves the passage for a given query.
More precisely, we plot $n_{q\shortto d}-\rho(d,\sigma_q)$, such that higher bar indicates that the document was retrieved at a higher rank for that query.

An ideal ranked list should arrange the vertical bars from the highest (queries exposing the passage at top ranking positions) to the lowest with no gap in between.
Except for Passage 3, the Residual model retrieves more exposing queries and ranks them higher than the baseline Reverse model. This shift in exposing queries in the ranked lists indicates that the proposed learning algorithm can learn a new metric space that can recover more exposing queries.

For Passage 1 and Passage 2, the Residual model furthermore retrieves many exposing queries which are originally not within the 200 nearest neighbors under the reverse baseline model. This example illustrates clearly that the nearest neighbor problem and the reverse nearest neighbor problem are not symmetric.
%The Residual model appears to rearrange the ranked list so that exposing queries retrieving the passages at higher ranks appear in the top positions.
%This shift in exposing queries in the ranked lists indicates that our proposed learning algorithm can effectively transform the original embedding space to an approximate reverse kNN embedding space.

Visualization of Passages 3 and 4 illustrates that, when a passage has fewer exposing queries overall, the RELQ metric tends to be more sensitive to the ranks $\rho(q,\psi_d)$ of highly exposing queries. This observations may, however, change under different user behavior assumptions for the RELQ metric. We analyze this aspect in the Appendix~\ref{subsec:user_model}.

\subsection{Impact of the training data size}\label{subsec:sample_size}
Approximate EQI trades exactness for performance; in the context of dual-encoder models, performance depends on the ability to train with little data. We examine the impact of training data size on model performance.
Table~\ref{tab:train_size} shows the results for the Residual model trained for $1000$ epochs.
We report $RELQ_{RBP,RBP}$ with $\gamma_{q \shortto d}=0.5, \gamma_{d \shortto q}=0.9$.
We find that increasing the number of both queries and training passages leads to diminishing returns. One-tailed paired t-tests are conducted along each column and each row. Except for the case, where the number of training queries is fixed at 500K and the number of training passages is increased from 400K to 600K, all increases in performance are statistically significant ($p < 0.01$) after Bonferroni correction. The Residual model trained on even a relatively small training data---$50$K queries and $200$K passages---shows good performance.

\begin{table}
\small
  \begin{center}
  \caption{Impact of the training sample size. Residual model performs better when either the sample size of training queries or training passages increases.}
    \label{tab:train_size}
\begin{tabular}{l|l|c|c|c}
\hline
& & \multicolumn{3}{c}{Passages} \\
\hline
& & $200,000$ & $400,000$ & $600,000$ \\
\hline
\multirow{3}[0]{*}{\rotatebox{90}{Queries}} & $50,000$ & 0.834 & 0.854 & 0.859
 \\
 & $250,000$ & 0.872 & 0.877 & 0.880
 \\
 & $500,000$ & 0.879 & 0.882 & 0.883
 \\
\hline
\end{tabular}
\end{center}
\end{table} 

\begin{figure}
    \centering
    \includegraphics[width=0.32\textwidth]{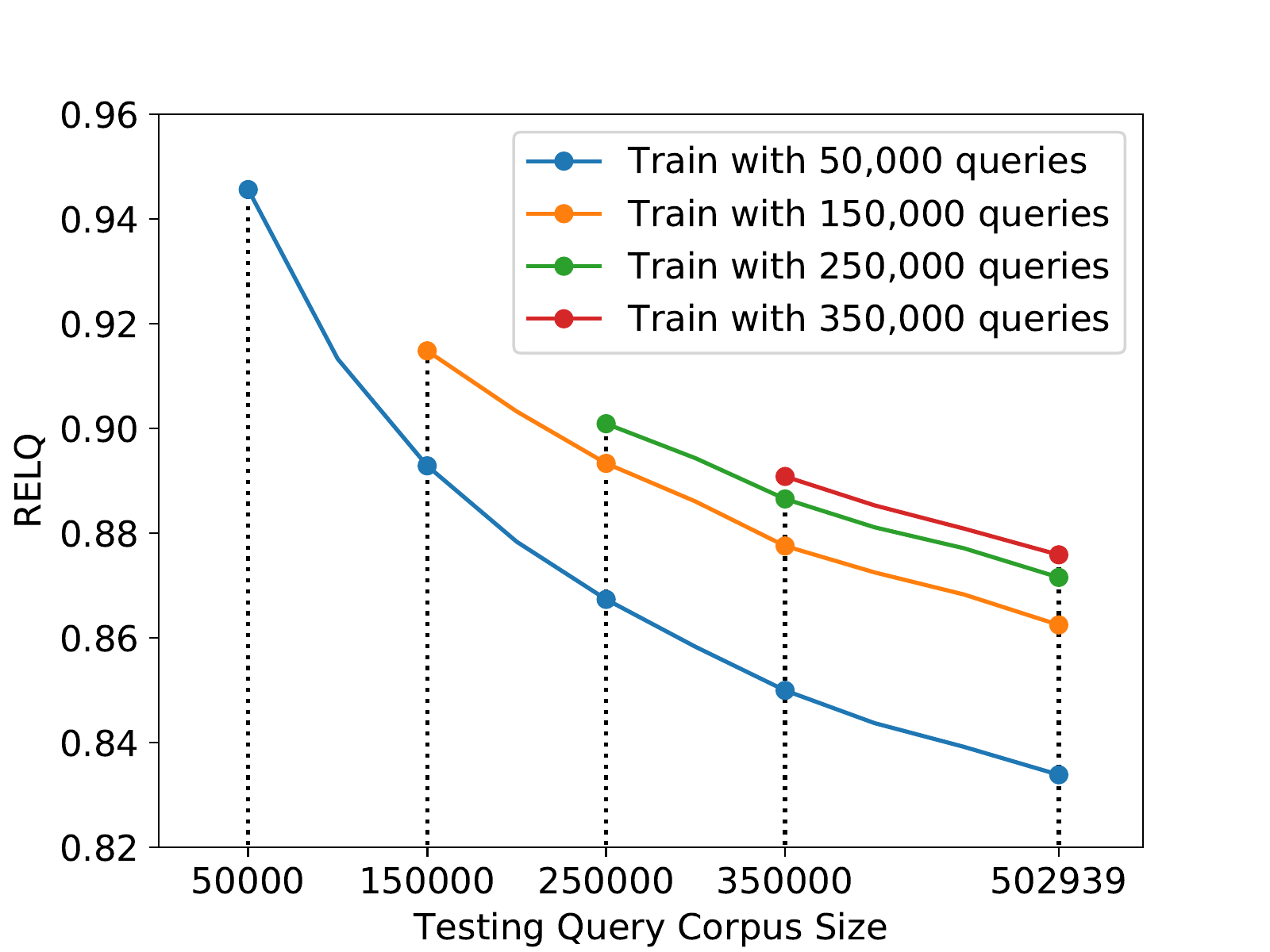}
    \caption{Each line is generated from a Residual model trained on a particular size of training queries. The starting point of each line is to train and test on the same query log. Along the line, query log is gradually expanding.}
    \label{fig:qset_evol}
\end{figure}
\vspace{.3em}

\begin{figure*}
    \centering
    \begin{subfigure}{0.32\textwidth}
    \centering
    \includegraphics[width=\textwidth]{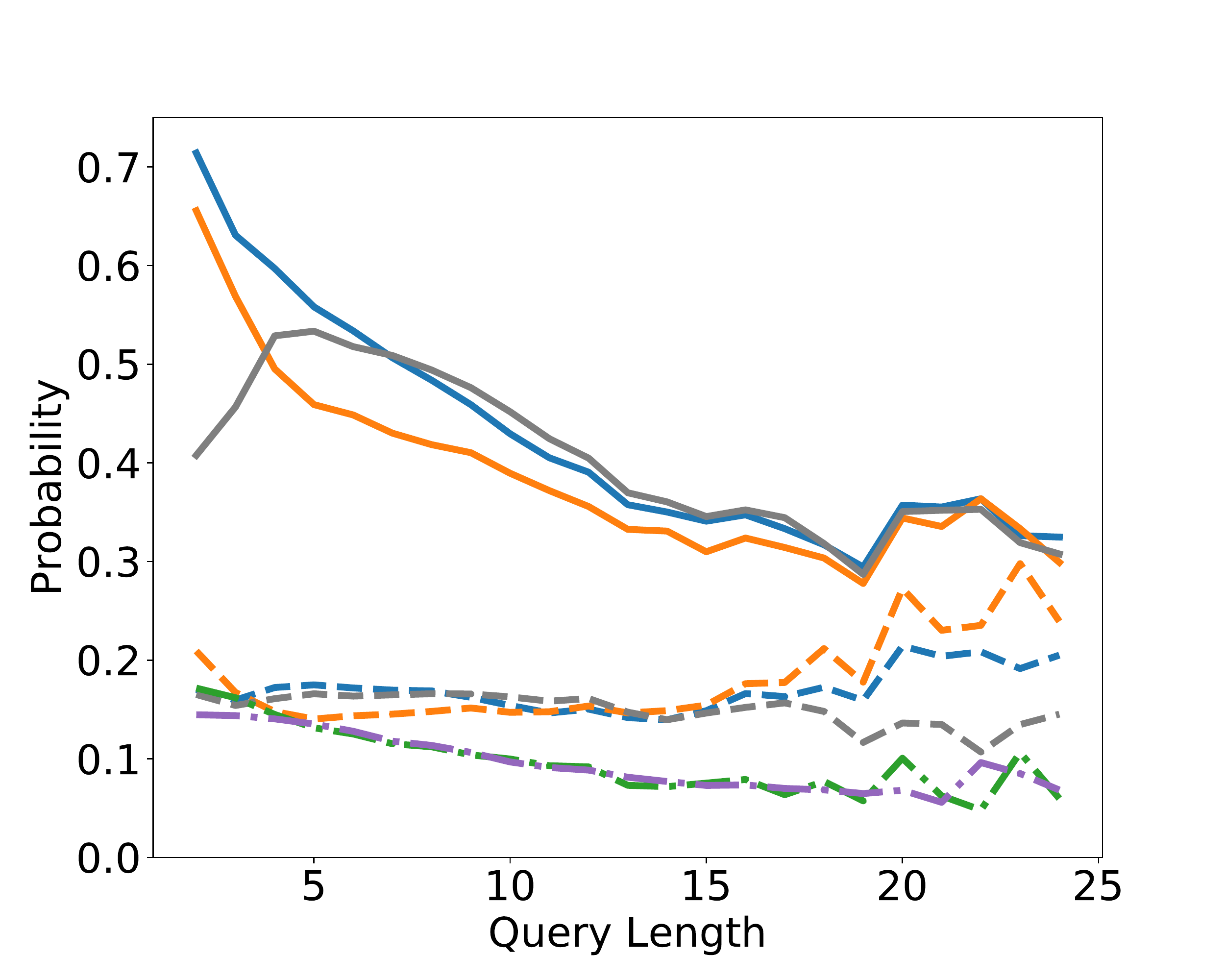}
    \caption{By query length}
    \label{fig:prob-length}
    \end{subfigure}
    \hfill
    \begin{subfigure}{0.30\textwidth}
    \centering
    \includegraphics[width=\textwidth]{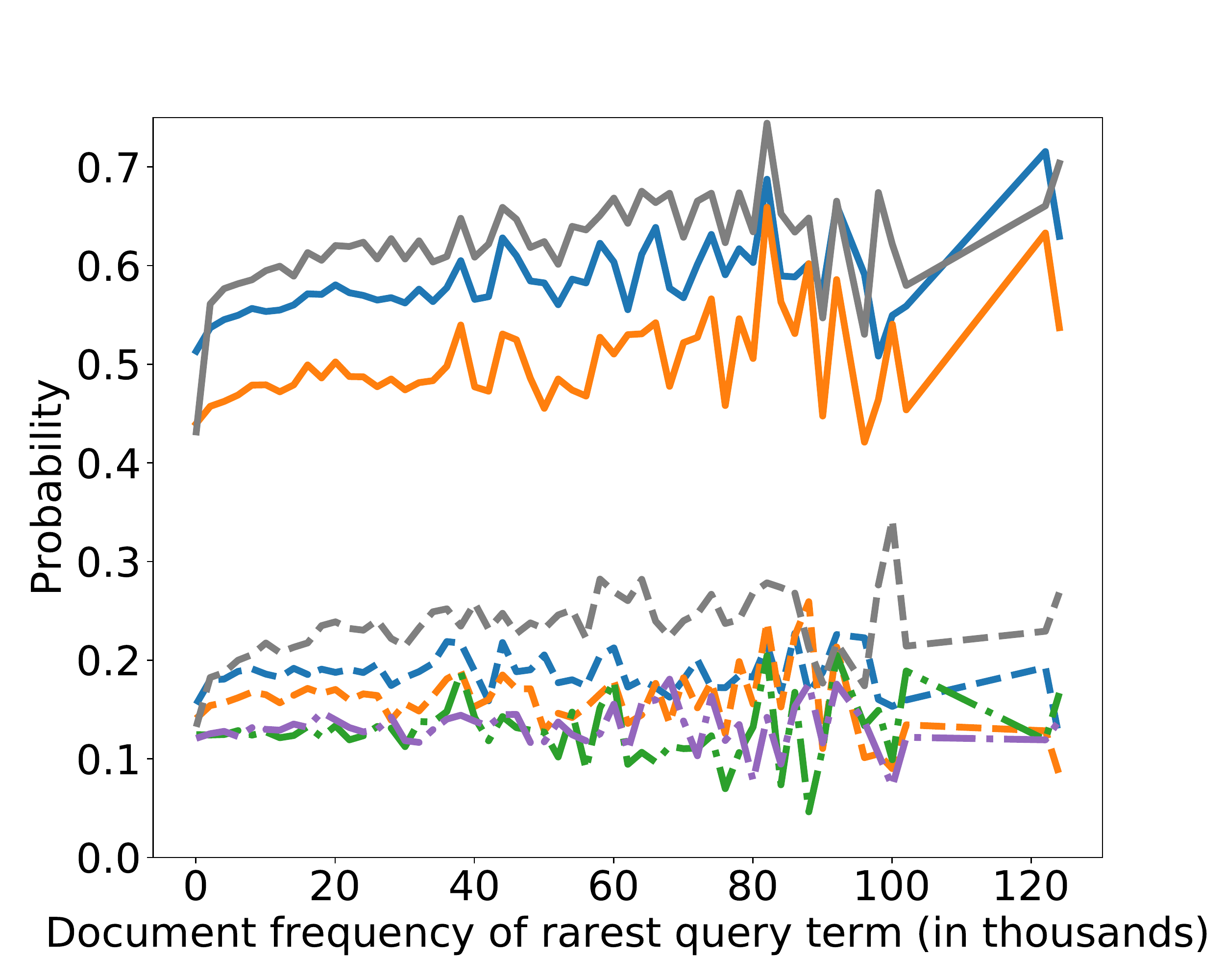}
    \caption{By document frequency of rarest query term}
    \label{fig:prob-df}
    \end{subfigure}
    \hfill
    \begin{subfigure}{.24\textwidth}
    \centering
    \includegraphics[width=\textwidth]{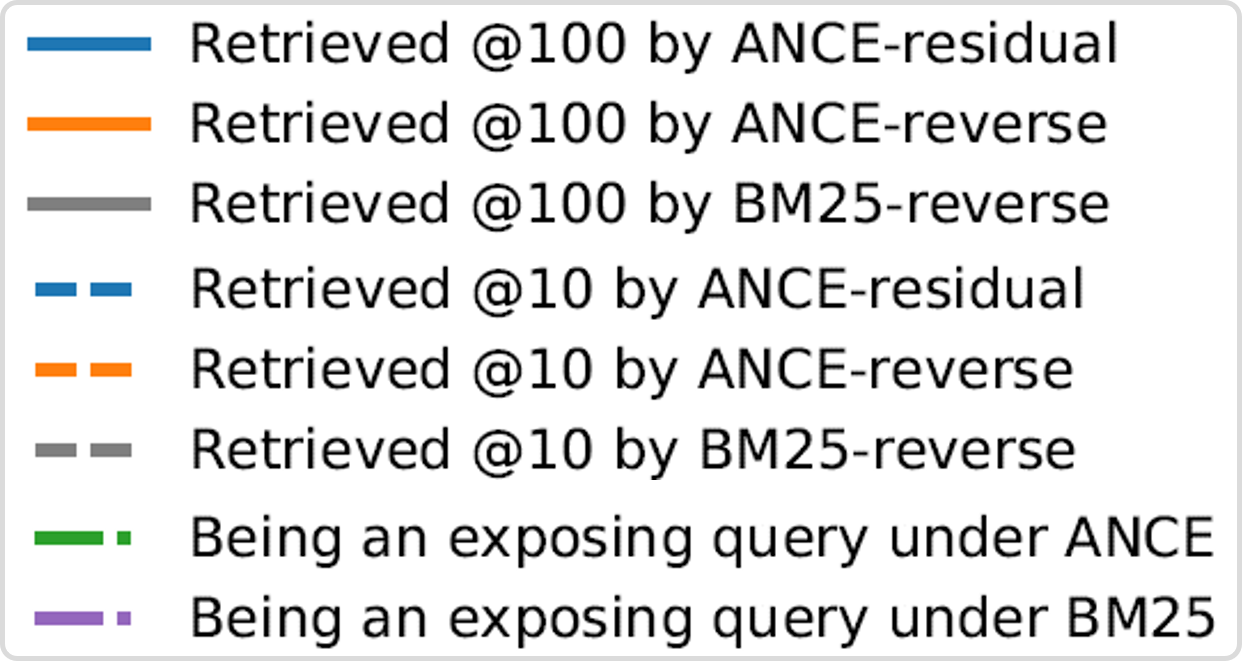}
    \end{subfigure}
\caption{Probability of being an exposing query for one of the test passages vs probability of retrieval.}
\label{fig:prob}
\end{figure*}

\begin{figure}
    \centering
    \tiny
    \includegraphics[width=0.42\textwidth]{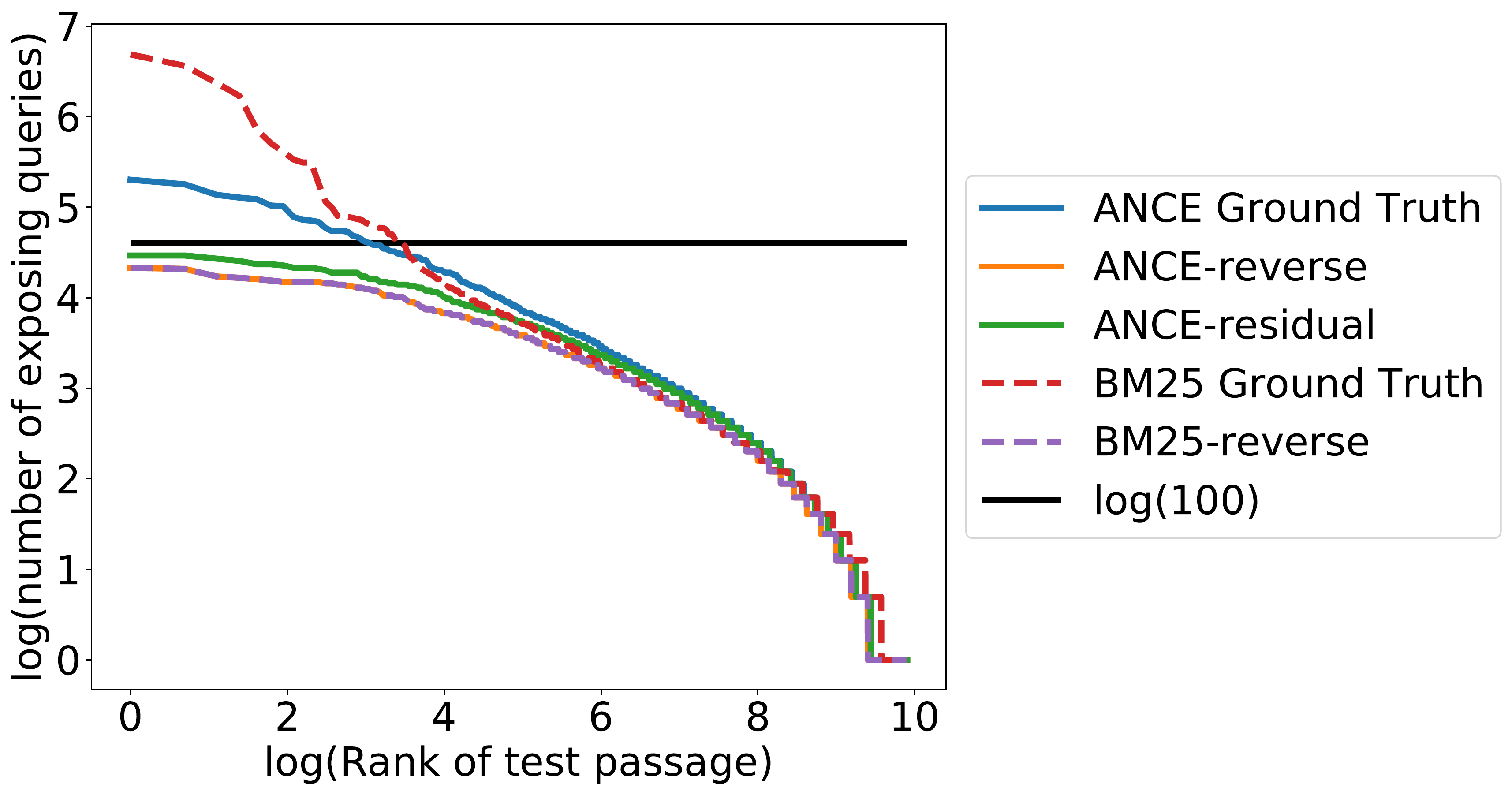}
    \caption{Rank-size distribution of the number of ground truth and retrieved exposing queries per passage. A few passages have a large number of exposing queries but there is also a long tail of passages with few or no exposing queries.}
    \label{fig:log_log}
\end{figure}

\subsection{Impact of expanding query collections}\label{subsec:expand_query}

The search ecosystem is dynamic. Searchers may issue previously unknown queries, while content producers may create new documents.
Note that the evaluation in Section~\ref{sec:model_performance} already covers the scenario in which exposure needs to be computed for new documents, as none of the test passages were seen or present in the corpus during training. We additionally examine the effects of a growing query collection. 

Our goal is to see how long the EQI model can sustain good retrieval performance without retraining as the query collection grows. This process is simulated by first sampling an initial set of queries available during training, and then test using a gradually expanding collection by embedding new queries without model retraining. The number of training passages is fixed as 200,000. Four Residual models are then trained for 1000 epochs on different sizes of a starting query collections: 50,000, 150,000, 250,000 and 350,000.
Note that this experimental protocol implies that new queries are drawn from the same distribution as existing queries.
Simulation results are illustrated in Figure \ref{fig:qset_evol}.
We report $RELQ_{RBP,RBP}$ with $\gamma_{q\shortto d}=0.5, \gamma_{d\shortto q}=0.9$.

\vspace{.5em}
\noindent\textbf{Decreasing concave trend along each line:} The decrease in model performance generally signals the need for model retraining.
In different real-world applications, service providers may have different tolerance levels to performance loss vs the retraining costs.
A visualization like the one in Figure \ref{fig:qset_evol} can be used as a monitoring tool for deciding when to retrain as the query collection expands.

The concave shape indicates that the model performance stabilizes. Recall that ANCE-Reverse achieves a RELQ of only 0.685 on the full query collection.
We reckon that the Residual model might eventually converge to a performance level still outperforming the Baseline model should the query collection increase further.

\vspace{.5em}
\noindent \textbf{Decreasing rate drop with larger training sample size:} The performance curves are flatter for models initially trained with more queries.
These models capture the characteristics of a fully expanded query collection better.

\vspace{.5em}
\noindent \textbf{Increasing task difficulty:} Both the decreasing trend along each curve and the decreasing starting value for each curve might suggest that the difficulty of the EQI task increases with the size of the query collection.
Interestingly, an opposite observation has been made about document retrieval systems by~\citet{hawking2003collection} who showed that the document retrieval precision increases with increasing collection size due to the sparsity of relevant documents in sub samples.
Reconciling this difference requires further analysis.
We hypothesize that the difference may be due to the fact that \citet{hawking2003collection} performed their analysis on term-matching-based retrieval models, whereas recent evidence shows that in the context of neural models an increase in the collection size may lead to a decrease in retrieval effectiveness~\citep{zamani2018neural}.

% \vspace{.5em}
% \noindent \textbf{Diminishing returns of retraining with bigger samples:} This observation is consistent with the results obtained in Section \ref{subsec:sample_size}. The benefits of retraining the model on a bigger query sample diminish as the sample size grows.

\subsection{Impact of query characteristics}
\label{subsec:query_analysis}

\vspace{.5em}
\noindent \textbf{Probability of retrieval vs. query length:}
We investigate whether the EQI models are more likely to retrieve longer queries.
As shown in Figure \ref{fig:prob-length}, under both BM25 and ANCE, the probability of being an exposing query for one of the test passages is slightly higher in case of shorter queries, \ie, queries with fewer than 10 words.
As for the top-10 retrieval probability, we find that both ANCE-reverse and ANCE-residual tend to slightly under-retrieve shorter queries and over-retrieve longer queries, while BM25-reverse retrieves equally regardless of query length. However, for the top-$100$ retrieval probability, we observe that all models are likely to significantly over-retrieve queries, in particular shorter ones.
% This indicates that as we go down the ranked list of exposing queries retrieved by either of these two models we are likely to uncover many short queries at lower ranks.

\vspace{.5em}
\noindent \textbf{Probability of retrieval vs. query term frequency:}
We investigate whether the models are able to retrieve both rare and common queries.
We measure the rarity of a query by computing the minimum document frequency of the query terms. This analysis sheds a light on the usefulness of the approach in providing privacy-awareness, as rare queries can contain personally identifying information.
Figure~\ref{fig:prob-df} demonstrates that ANCE-reverse, ANCE-residual and BM25-reverse tend to retrieve such queries at a rate roughly proportional to their probability of being exposing for one of the test passages.

\vspace{.5em}
\noindent\textbf{Distributions of exposing queries per passage:}
Figure \ref{fig:log_log} shows a log-log plot of the distribution of the number of exposing queries (y-axis) per passage. 
For all the methods, we find that there is a long tail of passages with very few or no exposing queries, while some passages may have tens, or even hundreds, of exposing queries. This result suggests the strong retrievability bias~\cite{azzopardi2008retrievability}, and is in line with prior work on search exposure~\cite{biega2017learning}.

\subsection{Discussion}
The results reported earlier in this section indicate that reversing the role of queries and documents can be a reasonable lightweight strategy for EQI, in particular when the model parameters are further optimized for the query retrieval task in the form of metric learning.
However, deploying an EQI system in the wild requires several other practical considerations.
While Section~\ref{subsec:sample_size} tells us that we can achieve reasonable effectiveness in exposing query retrieval using moderately sized training datasets (~50K queries), the required amount of training will depend on the size of the query corpus from which we are retrieving, as seen in Section~\ref{subsec:expand_query}.
As we can produce almost infinitely large training data for the query retrieval model automatically using the document retrieval model, the size of required training dataset itself may not be a big concern but may still have implications for training time of the query retrieval model and thus the attractiveness of EQI as a lightweight alternative to brute force aggregation.
% Another consideration is how the query retrieval performance changes with underlying shifts in query and passage distributions 
% which may have implications for how often the EQI model should be retrained.
% Due to space limitations, we leave that analysis outside the scope of this paper but hope to study in future work.

Another important concern may be around potential systemic biases in retrieved queries.
In Section~\ref{subsec:query_analysis}, we notice that the methods we study retrieve queries roughly in proportion to their probability of being exposing queries independent of query term rarity which is promising in terms of our ability to retrieve tail queries, and thus downstream privacy applications.
However, these models disproportionately over-retrieve shorter queries which may have implications for downstream applications such as quantifying the retrievability bias, or enforcing the right to be forgotten. Studying these effects is an interesting direction for future work.
% There may also be biases along other dimensions that are not studied here but which may be critical to the context in which these systems may be deployed.

% Finally, our optimization objective for the query retrieval model does not consider the choice of $\gamma_{q \shortto d}$ and $\gamma_{d \shortto q}$ which may result in lower effectiveness.
% In future work, listwise extension of our proposed loss functions could be explored to address this shortcoming.

%% file: 09-conclusion.tex
\section{Conclusions}
\label{sec:conclusion}

This paper focused on the search transparency problem of exposing query identification, that is, finding queries that expose documents in the top-ranking results of a search system. We studied the feasibility of approximate EQI as a retrieval problem for two classes of search systems (dual-encoder and traditional BM25 systems). We further demonstrated the potential of improving baseline EQI approaches through metric learning. 
We also derived an evaluation metric called Ranked Exposure List Quality (RELQ) that jointly models the behavior of two users who influence the EQI outcomes: a user of the EQI system and the document retrieval user. 
% The algorithm transforms the original document retrieval embedding space such that the nearest neighbor search in the transformed embedding space approximate the reverse nearest neighbor search in the original embedding space. 
% The learning algorithm includes an efficient training data generation algorithm that requires little computation cost on a small proportion of query and document collections. This design ensures our model competency in efficiency. Two simple but effective network architectures, Append and Residual, are proposed to shed lights on how deep neural network can effectively transform the embedding space to approximate the reverse nearest neighbor search task. 

Many promising future directions remain to be explored. Examples include developing better network architectures, algorithms for EQI in black-box search systems or for document retrieval models that use external click signals for ranking, developing generative models for query collections better grounded in downstream exposure tasks, as well as domain-specific quantitative and qualitative application and user studies. Our work provides a conceptual and empirical foundation for further study of EQI, which can contribute to advances in various areas of search and ranking: bias and fairness, privacy, data protection, security, and SEO.

% In Section \ref{sec:result}, we provide thorough analysis on model performance and retrieval results, which can be served as reliable reference for the potential users of our exposing query identification system. We experimentally prove our model robustness under different user behavior models and model generalizability to expanding query and document collections. Also, we show that our model is fair with respect to query length and query rarity, which is a desirable property in many downstream tasks. Our work in exposing query identification call for future work in further exploring following research questions. First, the possibility of expanding our metric learning algorithm to other search systems. Second, further works can be focused on incorporating generative NLP models to produce user specific candidates query set. Third, examine our model performance on other benchmark dataset, especially the dataset with clear user context or privacy concern.

%% file: 10-appendix.tex
\section{APPENDIX}

\begin{figure*}[th!]
    \centering
    \begin{subfigure}{0.99\columnwidth}
    \includegraphics[width=\textwidth]{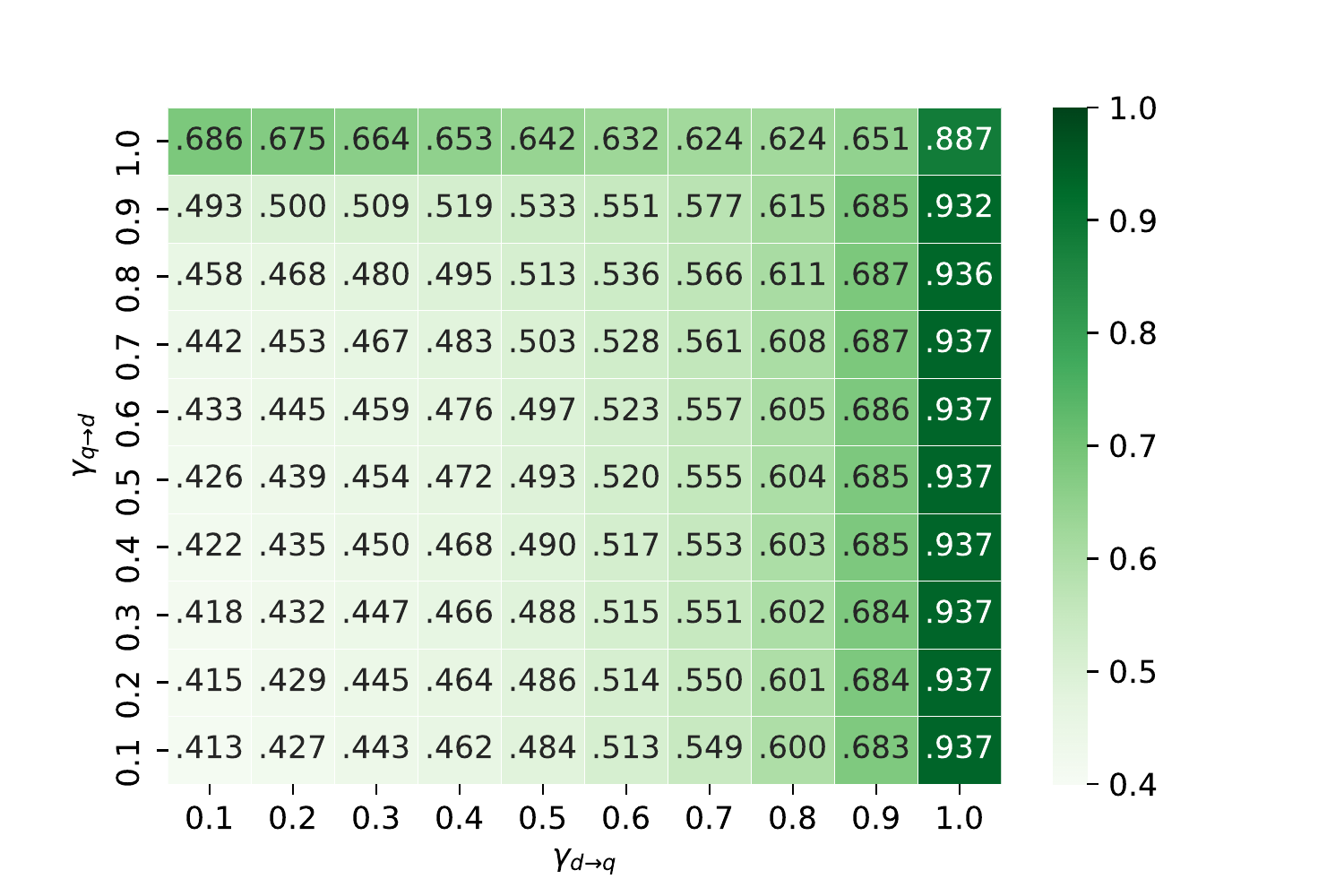}
    \caption{ANCE-reverse}
    \label{fig:baseline_heatmap}
    \end{subfigure}
    \begin{subfigure}{0.99\columnwidth}
    \includegraphics[width=\textwidth]{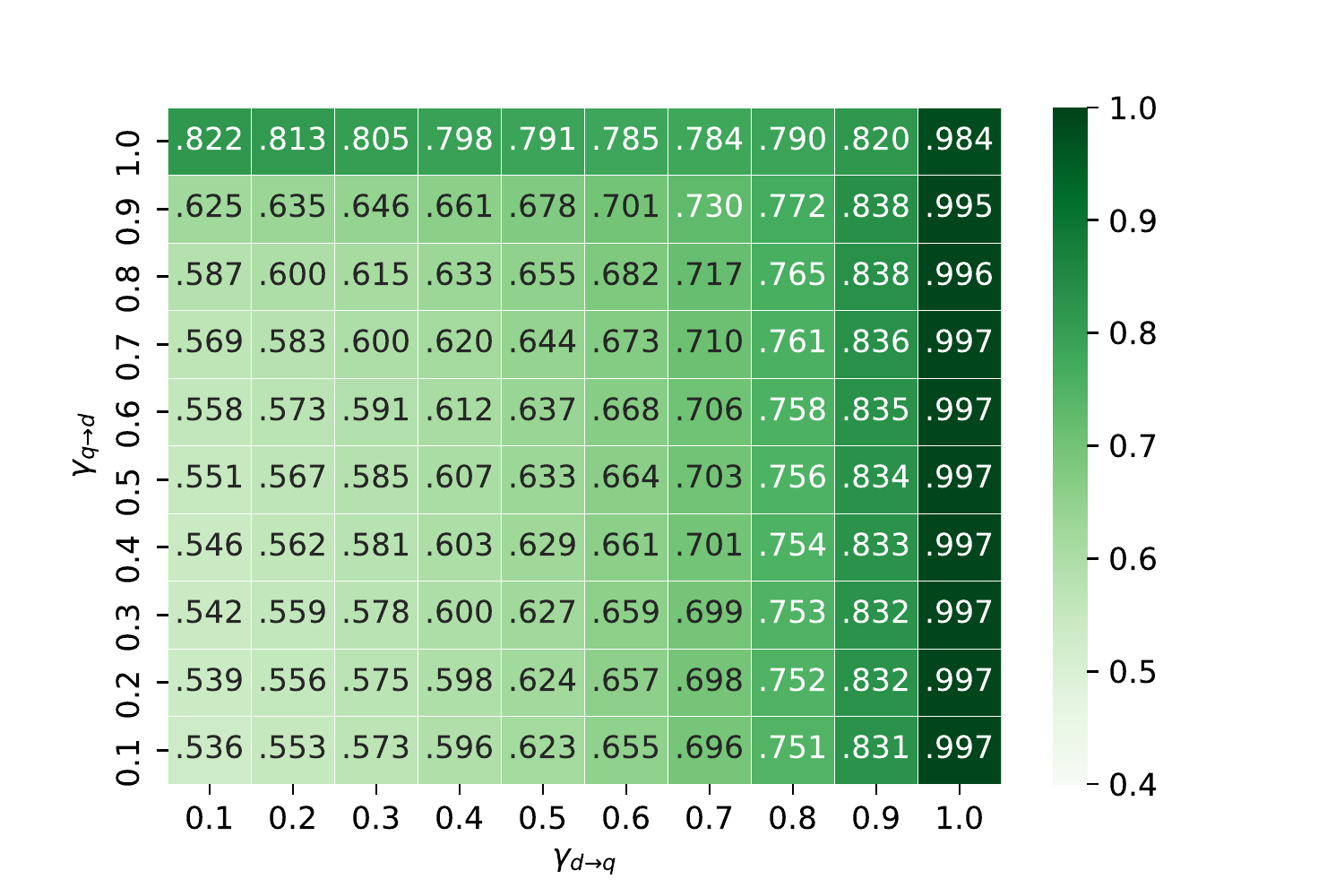}
    \caption{ANCE-residual}
    \label{fig:model_heatmap}
    \end{subfigure}
\caption{Model performance (RELQ values) for different user behavior models (the EQI system user on the x-axis, and the document retrieval user on the y-axis). The darker the color, the better the performance.}
\label{fig:heatmap}
\end{figure*}

\subsection{Impact of metric parametrization}
\label{subsec:user_model}

Under different user behavior assumptions, the same exposing query list will achieve different absolute value of RELQ.
We explore the relationship between user behavior and model performance for different combinations of user patience parameters $\gamma_{q\shortto d}$ (document retrieval user) and $\gamma_{d\shortto q}$ (EQI user).
Figure~\ref{fig:heatmap} presents the resulting performance heatmaps for ANCE-reverse and ANCE-residual.
Both heatmaps exhibit similar patterns.
% For a fixed EQI user model, the absolute value of RELQ is higher when the document retrieval users are more patient.
% This results suggests that finding exposing queries for less patient EQI users is more challenging.
Higher $\gamma_{q \shortto d}$ generally correlates with a high value of RELQ, except for very high values of $\gamma_{d \shortto q}$, such as $0.9$ or $1.0$, where a higher value of RELQ is observed for lower values of $\gamma_{q \shortto d}$.

These patterns are expected. Having a more patient user of the document retrieval system (high $\gamma_{q \shortto d}$) will mean more exposing queries per document: as the user goes deeper down the rank list more documents will get exposed to the query. Intuitively then, when the user of the EQI system is impatient, the presence of many ground truth exposing queries makes the task easier. 
However, when the EQI user is very patient and inspects the exposure set exhaustively, fewer ground truth exposing queries (impatient document retrieval user) do not lead to a higher task difficulty, and in this case, we indeed observe the highest absolute RELQ values.

ANCE-residual outperforms the ANCE-reverse model under all metric parametrizations.
Notably, however, ANCE-reverse performs very well for patient EQI users ($\gamma_{d \shortto q}=1.0$).
This result suggests that even though both models are able to retrieve many exposing queries, the ANCE-residual model tends to rank the exposing queries higher.